\documentstyle[11pt,newpasp,twoside,amsmath]{article}
\begin{document}
\title{Modeling of multicomponent radiatively driven stellar winds using
a Newton-Raphson method}
 \author{Ji\v r\'\i\ Krti\v cka}
\affil{\'UTFA P\v{r}F~MU,
            Kotl\'a\v{r}sk\'a 2, CZ-611~37~Brno, Czech~Republic\\
       Astronomick\'y \'ustav, AV \v{C}R, CZ-251~65~Ond\v{r}ejov, Czech~Republic}

\begin{abstract}
We present a simple method for the solution of one-compo\-nent and multicomponent
hydrodynamic equations based on the Newton-Raphson method. We show
that this method can be used for the solution  of stationary hydrodynamic
equations. This method has been used for the calculation of the low density 
stellar
wind models for which the multicomponent nature of the wind influences
the overall wind structure.
\end{abstract}

\section{Introduction}

Radiatively driven stellar winds are accelerated by the absorption of radiation
mainly in the resonance lines of such elements like C, N, O or Fe (see Castor,
Abbott, \& Klein 1975, hereafter CAK; Abbott 1982). These wind components have
much lower density than the stellar wind itself, which is composed mainly by
hydrogen and helium. 
The momentum obtained by the low-density absorbing wind component is transfered
to the high-density nonabsorbing wind component via
collisions between electrically charged particles 
(Springmann  \& Pauldrach 1992; Babel 1995). Clearly, multicomponent
radiatively driven winds of hot stars have a multicomponent nature. This
multicomponent nature is important mainly for the low-density stellar winds, which
shall be described by the multicomponent hydrodynamic equations.

Krti\v cka \& Kub\'at (2000; 2001a; 2001b) calculated
models of such multicomponent stellar winds using Newton-Raphson method. In this
paper we describe in detail method of solution of these multicomponent
hydrodynamic equations.

\section{Model equations}

Each wind component is described by a density $\rho_a$, radial velocity
${v_r}_a$, temperature $T_a$ and electrical charge $q_a$. Typically, we take
into account only three wind components (ie.~absorbing ions, nonabsorbing ions
and electrons).

Stationary spherically symmetric multicomponent radiatively driven stellar winds
are described by the continuity equation  for each component $a$ of the wind
\begin{equation}
\label{kontrov}
\frac{{\mathrm d} }{{\mathrm d}  r}\left({r^2\rho_a{v_r}_a}\right)  =  0,
\end{equation}
(note, that in the case of continuity equation of electrons a
right-hand term occurs due to the change of the wind ionization in the flow, see
Krti\v cka \& Kub\'at, 2001b),
momentum and energy equations for each component $a$
\begin{align}
\label{pohrov}
{v_r}_a\frac{{\mathrm d} {v_r}_a}{{\mathrm d} r}=
{g}_{a}^{\mathrm{rad}}-g-\frac{1}{{\rho}_a}\frac{{\mathrm d} }{{\mathrm d} r}\left({{a}_a^2{\rho}_
a}\right)
 +\frac{q_a}{m_a}E +
\sum_{b\neq a} g^{\mathrm{fric}}_{ab},\\
\label{enrov}
\frac{3}{2}{v_r}_a{\rho_a}\frac{{\mathrm d} a^2_a}{{\mathrm d} r}+
\frac{a_a^2\rho_a}{r^2}\frac{{\mathrm d} }{{\mathrm d} r}\left({r^2
{v_r}_a}\right)=
Q_a^{\mathrm{rad}}+\sum_{b\neq a} Q^{\mathrm{ex}}_{ab}+\sum_{b\neq a}
Q^{\mathrm{fric}}_{ab},
\end{align}
where $a_a$ is  isothermal sound speed, $E$ is charge
separation electric field, $g$ is gravitational acceleration, radiative
acceleration ${g}_{a}^{\mathrm{rad}}$ acts on free electrons and absorbing ions.
Radiative acceleration acting on absorbing ions is taken
in the CAK approximation with the finite disk
correction factor 
(Pauldrach, Puls, \& Kudritzki 1986)
and with force multipliers after Abbott (1982), slightly modified for the
multicomponent case (Krti\v cka \& Kub\'at 2000).
Frictional acceleration $g^{\mathrm{fric}}_{ab}$, frictional heating
$Q^{\mathrm{fric}}_{ab}$ and exchange of the heat $Q^{\mathrm{ex}}_{ab}$
 between wind components $a$
and $b$ are given by Burgers (1969; see also Krti\v cka \& Kub\'at 2001b). 

The only transitions which contribute to the radiative heating/cooling term
$Q_a^{\mathrm{rad}}$ in the static atmosphere are bound-free and free-free
transitions (Kub\' at, Puls, \& Pauldrach 1999; see also Kub\' at, this volume).
Because these transitions
deposit energy directly to electrons, this radiative heating/cooling
term is considered in electron energy equation.
Mean intensity $J_{\nu}$ at the base of the wind is taken as an emergent
radiation from a spherically symmetric static hydrogen model 
atmosphere (Kub\'at 2001).
 In the case of moving media
additional heating/cooling term occurs,
so-called Gayley-Owocki heating, which is caused by the 
dependence of a radiative force on a velocity via Doppler effect (Gayley \& Owocki
1994; Krti\v{c}ka \& Kub\' at 2001b).

The electrical charge $q_a$ is obtained using the approximate
ionization equilibrium described by Mihalas (1978, eq. [5-46]).

The system of hydrodynamic equations (\ref{kontrov}, \ref{pohrov}, \ref{enrov})
is closed using an equation for polarisation electric field in the form of
\begin{equation}
\label{polrov}
\frac{1}{r^2}\frac{{\mathrm d}}{{\mathrm d} r}\left({r^2
E}\right)=4\pi\rho=4\pi\sum_a\frac{q_a\rho_a}{m_a}.
\end{equation}

Finally, the system of equations solved is supplemented by 
appropriate boundary conditions. We start to calculate our models at the critical
point of nonabsorbing component. We also use zero current condition, the 
condition of
quasi-neutrality  and assume that the flow at the inner boundary is in
radiative
equilibrium and that the boundary temperature of all components is the
same. Boundary density of absorbing ions is determined using
abundance ratios.

\section{Method of solution}

Velocities, densities and other unknown variables are calculated at grid points.
We selected logarithmically spaced grid in radii with accumulation of grid points near
the stellar photosphere,
\begin{equation}
r_i=q r_{i-1}, \qquad 
q=\exp\left({\frac{\ln\frac{r_{\mathrm{N \! R}}}{r_1}}{\mathrm{N \! R}-1}}\right),
\end{equation}
where $i$ is a number of a given grid point and
$\mathrm{N \! R}$ is number of grid points. We use typically 50--300 grid points.

For the approximation of the derivative of expression $X$ we selected 
\begin{equation}
\label{aprder}
\left.\frac{{\mathrm d} X}{{\mathrm d} r}\right|_{r=r_i}\approx
\left\{
 \begin{array}{cc}
   {y}_i\frac{{X}_{i+1}-{X}_i}{\Delta {r}_{i+1}} +
            (1-{y}_i)\frac{{X}_i-{X}_{i-1}}{\Delta {r}_i}, & i<\mathrm{N \! R}, \\
   \frac{{X}_i-{X}_{i-1}}{\Delta {r}_i},& i=\mathrm{N \! R},
 \end{array} \right.
\end{equation}
where  
\begin{align}
\Delta {r}_i&={r}_i-{r}_{i-1},&
{\overline{r}}_i&=\frac{1}{2}\left({{r}_i+{r}_{i-1}}\right),&
{y}_i&=\frac{{r}_i-{\overline{r}}_i}{\overline{r}_{i+1}-{\overline{r}}_i}.
\end{align}
An appripriate approximation of the derivative is important,
numerical tests showed that approximation
(\ref{aprder}) gives the best convergence of the models. 

An approximation of hydrodynamical equations (\ref{kontrov}, \ref{pohrov},
\ref{enrov}) together with the equation for a polarisation electric field
(\ref{polrov}), equation of ionization equilibrium, and boundary conditions
can be formally written as
\begin{equation}
\label{vserov}
\mathsf{P}\boldsymbol\psi = 0,
\end{equation}
where the vector describing the solution has the form of
\begin{subequations}
\begin{gather}
\boldsymbol\psi=\left({\boldsymbol\psi_1,\boldsymbol\psi_2,\dots,\boldsymbol\psi_{\mathrm{N \! R}}}\right)^{\mathrm{T}},\\
\boldsymbol\psi_i=\left({\sum_a \rho_{a,i},\sum_a v_{ra,i},\sum_a T_{a,i},\sum_a z_{a,i},
E_i,\Delta v_{r,i}}\right)
\end{gather}
\end{subequations}
and $\mathsf{P}$ consists from all equations mentioned above.

For the solution of the nonlinear system of model equations (\ref{vserov}) we
selected Newton-Raphson method. The solution of Eqs.(\ref{vserov}) can be
obtained using iterative procedure in the form of
\begin{equation}
\label{syslin}
\mathsf{J}^{n}\delta\boldsymbol\psi^{n+1}=-\mathsf{P}^{n}\boldsymbol\psi^{n},
\end{equation}
where $\boldsymbol\psi^{n}$ denotes solution in the $n$-th iteration, 
$\delta\boldsymbol\psi^{n+1}$ is a correction of the solution and the
Jacobian is
\begin{equation}
{J}_{kl}^{n}=\frac{\partial P_{k}}{\partial\psi_l}.
\end{equation}
The Newton-Raphson method is similar to the well-known method of 
complete linearization (Auer \& Mihalas 1969).
The analytical form of Jacobi matrix ${J}_{kl}^{n}$ can be easily obtained from
Eqs.(\ref{vserov}) and for the case of three-component radiatively driven
stellar wind was calculated by Krti\v{c}ka (2001). However, the value of
corrections $\delta\boldsymbol\psi^{n+1}$ in each iterative step is limited 
by $0.7\boldsymbol\psi^{n}$ during the calculation of a model below the
critical point and by $0.1\boldsymbol\psi^{n}$ during the calculation
of a model above the critical point. The velocity differences between wind
components are  limited similarly.

Another problem for the calculation of wind models is the inclusion of critical
point condition. Critical point is a point where the wind velocity is equal to
the sound speed. Generally, there are two possibilities of inclusion of the
critical point condition. Nobili \& Turola (1988) showed that critical point
condition can be directly included into the set of linearised equations.
However, due to some numerical problems, we selected the so called "shooting
method". We change the density at the base of the wind in order to fulfil the
critical point condition. When the base density and model upstream the
critical point are obtained, we calculate model downstream the critical point.
The initial model for the Newton-Raphson iterations is selected in
such a way that it converges to the correct branch of CAK solution appropriate 
for the flow downstream the critical point (see CAK).

For the solution of the system of linearised equations (\ref{syslin}) we use
numerical package LAPACK. We also applied the Gaussian elimination, however we
found that the package LAPACK is faster and even more accurate.

Newton-Raphson method represents quite powerful tool for the solution of
stationary hydrodynamic equations. The iterations converge very fast (typical
relative change is of order $10^{-10}$ after 20 iterations), the number of
equations solved can be easily extended and it is possible to calculate
multidimensional models. On the other hand the model convergence 
depends on initial estimate and we should also check the
dynamical stability of obtained solution.

\begin{acknowledgements}
The author is greatly indebted to Dr.~J.~Kub\'at for discussion and cooperation on this topic.
This work was supported by grants GA \v{C}R 205/01/0656, 205/02/0445,
and by projects K2043105 and AV0Z1003909.
\end{acknowledgements}

\end{document}